\documentclass[%
 preprint, 
 amsmath,amssymb,
 aps, physrev,
]{revtex4-2}

\usepackage{graphicx}
\usepackage{dcolumn}
\usepackage{bm}

\begin{document}

\preprint{AIP/123-QED}

\title{Quantitative correlation between structural (dis-)order and diffuseness of phase transition in lead scandium tantalate}

\author{T. Granzow$^1$}
\email{torsten.granzow@list.lu}

\author{A. Aravindhan$^{1,2}$}%
\author{Y. Nouchokgwe$^1$}%
\author{V. Kovacova$^1$}%
\author{S. Glinsek$^1$}%

\author{S. Hirose$^3$}
\author{T. Usui$^3$}

\author{H. Ur\v{s}i\v{c}$^4$}
\author{I. Gori\v{c}an$^4$}

\author{W. Jo$^5$}

\author{C.-H. Hong$^6$}

\author{E. Defay$^{1,2}$}%
 
\affiliation{%
$^1$ Smart Materials Unit, Luxembourg Institute of Science and Technology, 41 rue du Brill, 4422 Belvaux, Luxembourg}%

\affiliation{%
$^2$ University of Luxembourg, Dept. of Physics and Materials Science, Belvaux, Luxembourg}%

\affiliation{$^3$ Murata Manufacturing Co., Nagaokakyo, Kyoto, Japan}%

\affiliation{%
$^4$ Jo\v{z}ef Stefan Institute, Ljubljana, Slovenia}%
\affiliation{$^5$ School of Materials Science and Engineering, Ulsan National Institute of Science and Technology, Korea}%
\affiliation{$^6$ MLCC Manufacturing Technology Team, Samsung Electro-Mechanics, Korea}%
\date{\today}

\begin{abstract}
Ferroelectrics show a phase transition to a paraelectric phase at a well-defined transition temperature. Introducing disorder makes this transition diffuse, and the system becomes a relaxor. Since the degree of (dis-)order is usually manipulated by varying the chemical composition, it is difficult to establish a direct relationship between disorder and the degree of diffuseness. Perovskite structured lead scandium tantalate (Pb[Sc$_{1/2}$Ta$_{1/2}$]O$_3$, PST) offers the opportunity to tune the character of the transition by thermal annealing without changing the stoichiometry. Here it is demonstrated that there is a linear correlation between the structural ordering, quantified by the intensity ratio $S$ of the pseudocubic (111)/(200) x-ray diffraction peaks, and the diffuseness parameter $\gamma$ deduced from temperature-dependent dielectric spectroscopy. The relation is universal, independent of whether the sample is a thin film, multilayer capacitor or bulk ceramic, and also independent of the absolute value of the dielectric permittivity.
\end{abstract}


\maketitle

\section{Introduction}

Ferroelectric materials are characterized by a permanent electrical dipole moment in their crystallographic unit cells. Their long-range ordered interaction results in a macroscopic 'spontaneous' electrical polarization $P_S$ in the absence of an external electric field. The ferroelectric phase usually breaks down into domains, {\it i.e.}, regions of homogeneous polarization direction separated by domain walls. Ferroelectrics undergo a transition to a paraelectric high-temperature phase at a well-defined transition temperature $T_C$. For $T> T_C$, the dielectric permittivity $\varepsilon'$ typically follows the Curie-Weiss law:

\begin{equation}\label{CW}
    \frac{1}{\varepsilon'}=\frac{T-T_0}{C},
\end{equation}
where $T_0$ and $C$ are the Curie-Weiss temperature and constant, respectively. For a second-order transition, $T_0 = T_C$, while for a first-order transition $T_0 < T_C$.

In contrast, 'relaxor ferroelectrics' or 'relaxors' also carry a spontaneous polarization, but have a 'diffuse' transition that is accompanied by a broad maximum of $\varepsilon'(T)$ at a temperature $T_M$ that increases with increasing measurement frequency, but does not involve a macroscopic symmetry breaking \cite{Cross94}. Their dielectric response is influenced by local dipole moments below what is now called Burns temperature $T_B >> T_M$ \cite{Burns83}, though the usefulness of this assignment has been questioned \cite{Bobnar11}. These dipoles form fluctuating polar nanoregions (PNR) in the ergodic relaxor state. Upon cooling, the PNR grow, their dynamics slow down drastically and the systems enters a disordered non-ergodic relaxor state \cite{Tagantsev98}. In the ergodic state, high electric polarization and mechanical strain can be reversibly induced by an electric field, making relaxors attractive for applications such as capacitive energy storage \cite{Palneedi18} or high-strain piezoelectric actuators \cite{Jo12}. The possibility to manipulate the relaxor character of a system is therefore highly desirable.

Relaxor properties are attributed to structural disorder suppressing ferroelectric long-range order between local dipoles. A classical example is perovskite-structured Pb(Mg$_{1/3}$Nb$_{2/3}$)O$_3$ (PMN), where the perovskite B-site is randomly occupied by Mg$^{3+}$ and Nb$^{5+}$ ions. In recent years, a 'high-entropy' approach has gained in popularity \cite{Zhou22} to modify structural disorder by putting ions with different valence states or radii on the same crystallographic lattice position, with the aim of increasing relaxor features. To quantify the 'relaxor character' of a system, the 'diffuseness parameter' $\gamma$ was introduced by modifying eq.\,\ref{CW}: 

\begin{equation}\label{Diffuse}
    \frac{1}{\varepsilon'}-\frac{1}{\varepsilon_M'(f)}=\frac{1}{C}*(T-T_M(f))^\gamma,
\end{equation}
where $T_M'(f)$ and $\varepsilon_M'(f)$ are the temperature where the maximum of permittivity is observed at the measurement frequency f, and the value of permittivity at this temperature, respectively \cite{Uchino82}. For a ferroelectric following the Curie-Weiss law, $\gamma = 1$, while a value of $\gamma=2$ is expected for 'perfect' relaxors  \cite{Kirillov73}; any number between these values is considered to indicate a more or less 'relaxor-like' behavior. It has to be noted that while eq.\,\ref{Diffuse} is phenomenologically well established, its mathematical derivation relies on assumptions that have been shown to not accurately reflect the physics underlying the relaxor behavior \cite{Ahn16}. It remains an open question if this derivation can be based on a more accurate description of relaxor system such as the random field model \cite{Westphal92}. The connection between structural disorder and diffuseness of phase transition has been observed qualitatively \cite{Baskaran01}, but establishing a quantitative correlation has so far proven elusive when disorder is created by modification of sample stoichiometry \cite{Lei06}.

Selection of ions to achieve high entropy is dictated by crystal chemistry and cannot consider the influence on local dipole moments and thereby macroscopic polarization. Thus, the increase in relaxor properties by adding aliovalent ions usually comes at the price of reduced polarization. Among the very few systems where structural disorder and relaxor character can be modified without changing the stoichiometry are perovskite-structured Pb(Sc$_{1/2}$Ta$_{1/2}$)O$_3$ (PST) and its sister compound Pb(Sc$_{1/2}$Nb$_{1/2}$)O$_3$ (PSN). When quenched from high temperatures, Sc$^{3+}$ and Ta$^{5+}$ or Nb$^{5+}$, respectively, are randomly distributed on the perovskite B-site. This structural disorder makes the system a relaxor \cite{Stenger79}. When annealed and cooled slowly, the trivalent and pentavalent ions order in a Rocksalt structure, and the system is a ferroelectric with a first-order phase transition \cite{Stenger79}. The degree of structural (dis-)order can be determined by x-ray diffraction (XRD): for a fully disordered system, the pseudocubic (111)$_{\rm pc}$ reflex is symmetry forbidden, while in the Rocksalt-ordered system it appears as a superstructure reflex. Putting the XRD intensity $I_{111}$ of this peak in relation to that of a reflex that is symmetry allowed independent of the superstructure, such as the $I_{200}$ of the pseudocubic (200)$_{\rm pc}$ peak, gives the ordering coefficient $S$:

\begin{equation}
    S^2 = \frac{\left(\frac{I_{111}}{I_{200}}\right)}{\left(\frac{I_{111}}{I_{200}}\right)_{S=1}}.
    \label{Orderingeq}
\end{equation}

Here, $\left(\frac{I_{111}}{I_{200}}\right)_{S=1}$ is the peak intensity ratio for the perfectly ordered system; it is not determined experimentally, but calculated from the theoretical angular distribution of x-rays diffracted on a structure with full Rocksalt-ordering on the B-site. Literature gives a value of $1.36\pm0.04$ for this ratio \cite{Pietraszko04,Shebanov88}.

In this paper, we show that structural ordering parameter $S$ and phase transition diffuseness parameter $\gamma$ are strongly correlated in PST. This provides a quantitative relation between structural order and phase transition diffuseness that covers strongly different sample geometries and mechanical boundary conditions.

\section{Experimental details}

Using established processing routes described in detail in the cited references, different types of PST samples were prepared: thin films (TFs) with a thickness of 200\,nm were produced by spin coating on c-sapphire substrates and fitted with Pt interdigitated electrodes (IDEs) with a total of 50 pairs of fingers with an effective length of 370\,$\mu$m, finger width of 5\,$\mu$m and a gap width of 3\,$\mu$m by sputter deposition and lift-off photolithography \cite{Aravindhan24}. Multilayer capacitors (MLCs) with a total of 9 active layers with a layer thickness of 38\,$\mu$m and an active area of 49\,mm$^2$ per layer were fabricated by tape casting \cite{Torello20}, and disc-shaped bulk ceramic samples with a diameter of 8.5\,mm and a thickness of 1\,mm by conventional powder processing \cite{Aso19}. Since the as-prepared TF and bulk samples showed only a low degree of B-site occupation ordering, some of them underwent a thermal annealing step to increase ordering \cite{Aso19}. The crystallographic structure of all samples was investigated by XRD on a Bruker D8 Discover diffractometer using Cu k$_{\alpha}$ radiation. In the case of the MLCs and bulk PST, measurements were performed on powders obtained by crushing the samples and then kept at 150\,$^{\circ}$C for 30 minutes to release machining-induced mechanical stress. For the randomly oriented powders, normal $\theta-2\theta$ scans were performed. As the PST films showed a (100) preferential orientation, $\theta-2\theta$ scans were performed under a tilt angle of $\chi = 54.7^{\circ}$, the angle between the pseudocubic (111) and (200) planes, to observe the (111) plane under optimum conditions \cite{Aravindhan24}.

Dielectric permittivity was measured at frequencies between 1\,MHz and 1\,kHz using a Novocontrol Concept 40 Dielectric Spectrometer system. The temperature range was adjusted to the transition temperatures of the different systems. Measurements with different rates of temperature change showed that results became independent of the ramp rate below 2\,K/min; all measurements presented here were taken at effective rates of 1\,K/min or lower.

\section{Experimental results}

\begin{figure}[htbp]
\includegraphics[width=1.0\textwidth]{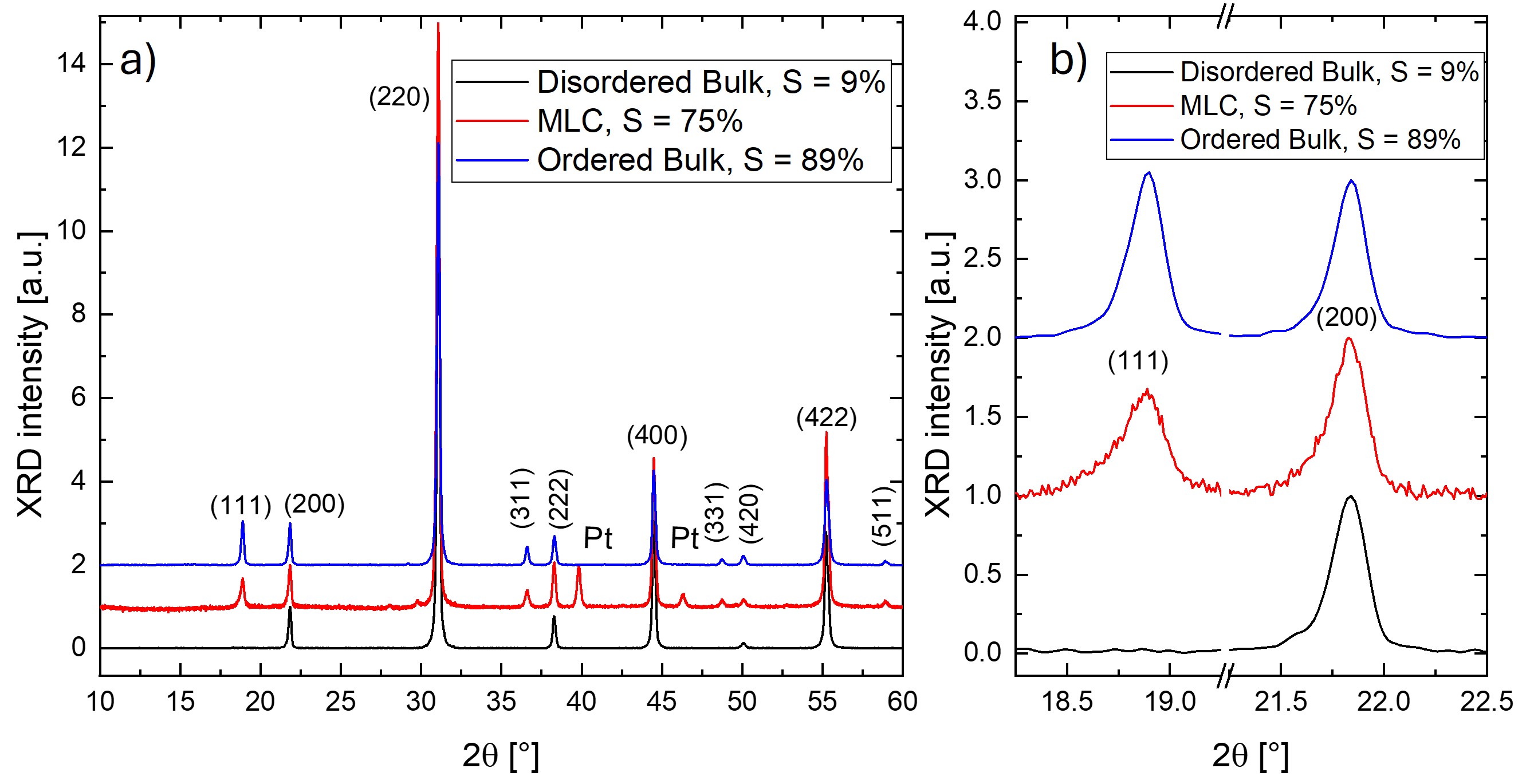}
\caption{\label{f1} (a) X-ray diffractograms of a crushed bulk sample with ('Ordered Bulk') and without ('Disordered Bulk') thermal anneal, respectively, and  the crushed MLC. (b) Close-up view of the region of the pseudocubic (111) superstructure peak and (200) regular lattice peak.}
\end{figure}

X-ray diffraction revealed for all samples a perovskite structure without secondary phases, except for reflections belonging to the metallic Pt of the internal electrodes in the crushed MLCs. There was no peak splitting that might indicate a distortion of the cubic phase. The diffractrograms of three samples - the crushed bulk samples without and with a thermal annealing treatment to increase B-site ordering as well as the MLC - are shown in Fig.\,\ref{f1}(a), normalized to the (200) peak height. To assess the degree of ordering, the (111) and (200) peaks are shown at a larger scale in Fig.\,\ref{f1}(b).

The (111) superstructure peak is notably larger in the XRD pattern obtained from the bulk sample where B-site ordering was increased by thermal annealing ('Ordered Bulk') than from the MLC; in the bulk sample that did not undergo this treatment ('Disordered Bulk'), the corresponding peak is hardly visible. Analyzing the peak areas according to eq.\,\ref{Orderingeq}, one obtains values of $S=0.09$ for the 'Disordered Bulk' sample, $S=0.89$ for the 'Ordered Bulk' sample and $S=0.75$ for the MLC. Similarly, the as-deposited thin film ('Disordered film') did not display any (111) peak and therefore $S=0$, while for a thermally annealed 'Ordered film' $S=0.35$ was reached. The highest degree of ordering was obtained with $S=0.98$, {\it i.e.} nearly perfectly ordered, for a second bulk sample annealed for a longer time.

\begin{figure}[htbp]
\includegraphics[width=1.0\textwidth]{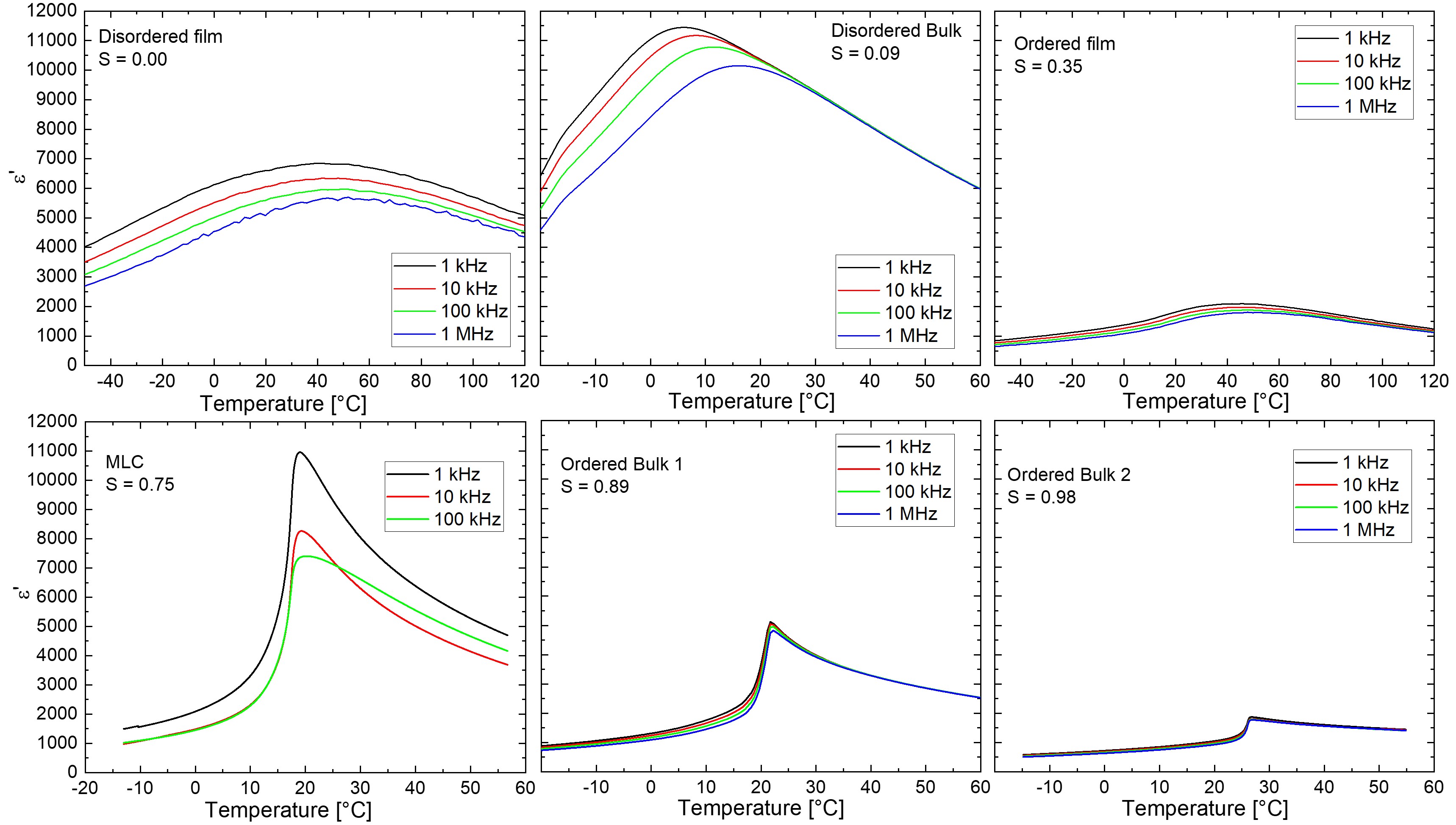}
\caption{\label{f2} Temperature-dependent permittivity $\varepsilon'$ for six PST samples with different geometries and different degrees of B-site ordering at frequencies between 1\,MHz and 1\,kHz.}
\end{figure}

The dielectric permittivity $\varepsilon'$ is presented as function of temperature in Fig.\,\ref{f2} at frequencies from 1\,kHz to 1\,MHz for all samples except for the MLCs, where the data at 1\,MHz was unreliable: an electrical resonance appeared at high frequencies due to the very large capacitance. The change in the behavior is obvious: with increasing B-site ordering, the peak denoting the transition becomes sharper and more pronounced. Table \ref{Tab1} gives the maximum value of permittivity $\varepsilon_M'$ and the temperature $T_M'$ where this maximum is observed for all samples and frequencies.

\begin{table}

\begin{tabular}{| l | c | c | c | c | c | c | c | c | c |}
\hline
Sample & S & \multicolumn{2}{l} {1 kHz} & \multicolumn{2}{l}{10 kHz} & \multicolumn{2}{l}{100 kHz} & \multicolumn{2}{l}{1 MHz} \\
\hline
  &   & $T_M$ [°C] & $\varepsilon_M'$ & $T_M$ [°C] & $\varepsilon_M'$ & $T_M$ [°C] & $\varepsilon_M'$ & $T_M$ [°C] & $\varepsilon_M'$ \\
\hline
Disordered film & 0.00 & 42 & 6830 & 45 & 6330 & 48 & 5950 & 52 & 5650 \\
\hline
Disordered bulk & 0.09 & 6 & 11400 & 8 & 11200 & 12 & 10800 & 16 & 10200 \\
\hline
Ordered film & 0.35 & 44.0 & 2090 & 45.4 & 1970 & 46.9 & 1880 & 49.7 & 1800 \\
\hline
MLC & 0.75 & 19.0 & 10950 & 19.3 & 8260 & 20.2 & 7420 & -  & -  \\
\hline
Ordered bulk 1 & 0.89 & 21.7 & 5120 & 21.7 & 5050 & 22.2 & 4960 & 22.1 & 4820 \\
\hline
Ordered bulk 2 & 0.98 & 26.8 & 1860 & 26.8 & 1800 & 26.9 & 1780 & 27.0 & 1770 \\
\hline

\end{tabular}
\caption{\label{Tab1} Temperature $T_M$ at which the maximum of permittivity $\varepsilon_M'$ is observed in all samples for different frequencies.}
\end{table}

\begin{figure}[htbp]
\includegraphics[width=1.0\textwidth]{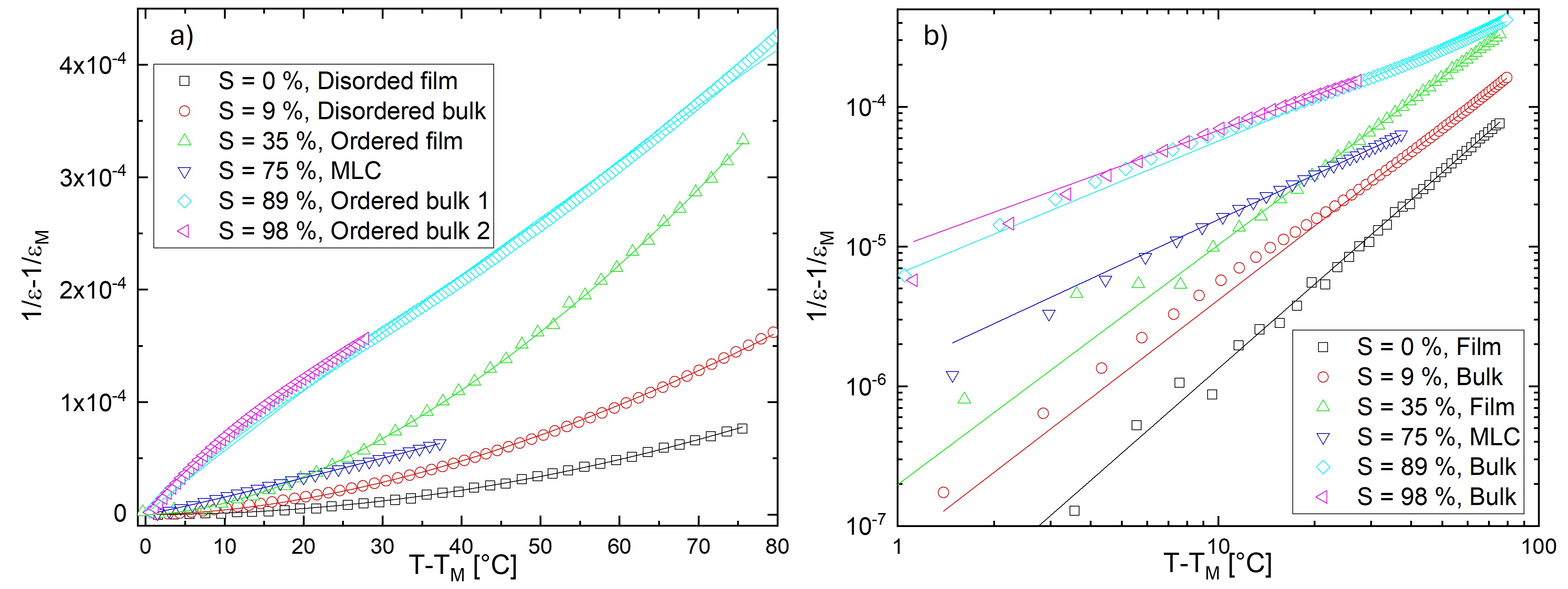}
\caption{\label{f3} $\frac{1}{\varepsilon'(T)}-\frac{1}{\varepsilon_M'}$ at 1\,kHz as a function of $T-T_M$ for $T\geq T_M$ (a) on a linear scale and (b) on a double log scale.}
\end{figure}

Using these values of $T_M$ and $\varepsilon_M'$, Fig.\,\ref{f3}(a) exemplarily shows $\frac{1}{\varepsilon'(T)}-\frac{1}{\varepsilon_M'}$ at 1\,kHz as a function of $T-T_M$ for $T\geq T_M$ on a linear scale, while Fig.\,\ref{f3}(b) gives the same data on a double logarithmic scale. The continuous lines in part (a) of the figure are fits of eq.\,\ref{Diffuse} to the data, while in part (b) the represent linear fits to the data on the logarithmic scale. In both cases, the fitted lines correspond well to the experimental data. 

\section{Discussion}

There are several trends visible in the permittivity of the different systems: first, the frequency dependence of $T_M$ decreases with increasing structural ordering, following the expectation that the systems transition from a relaxor behavior for low ordering to a ferroelectric behavior for high ordering: for the disordered film with $S=0$, there is a difference of 10\,$^{ \circ}$C between the $T_M$ at 1\,kHz and 1\,MHz, while for the highly ordered bulk sample with $S=0.98$, the spread of 0.1\,$^{ \circ}$C vanishes within the error margin. This observation holds across all sample geometries, despite the fact that the $T_M$ of the films is significantly higher than that of the bulk samples: while the bulk samples are unstressed, the films are subject to a high biaxial compressive stress induced by the clamping to the substrate, which stabilizes the polar structures and raises the transition temperatures.

The second general trend is a decrease in permittivity in each type of system as the ordering increases, in agreement with the assumption that highly mobile polar nanoregions that can easily follow the ac field progressively turn into ferroelectric domains, with only the domain walls reacting to the field excitation. This trend has been reported before for bulk PST \cite{Stenger79, Setter80}, but for thin films the opposite behavior has been described \cite{Brinkmann07} and explained by the formation of Pb- and O-vacancies during high-temperature film processing. The films prepared for the present work apparently do not suffer from these defects and therefore follow the same trend as the bulk samples. Overall, the dielectric response of the thin films is strongly reduced by the mechanical stress exerted by the substrate, an observation that was reported before \cite{Brinkmann07b}. This effect absent in the unclamped bulk samples as well as the MLCs, where the clamping of the thick layers by the electrodes is negligible.

\begin{figure}[htbp]
\includegraphics[width=0.5\textwidth]{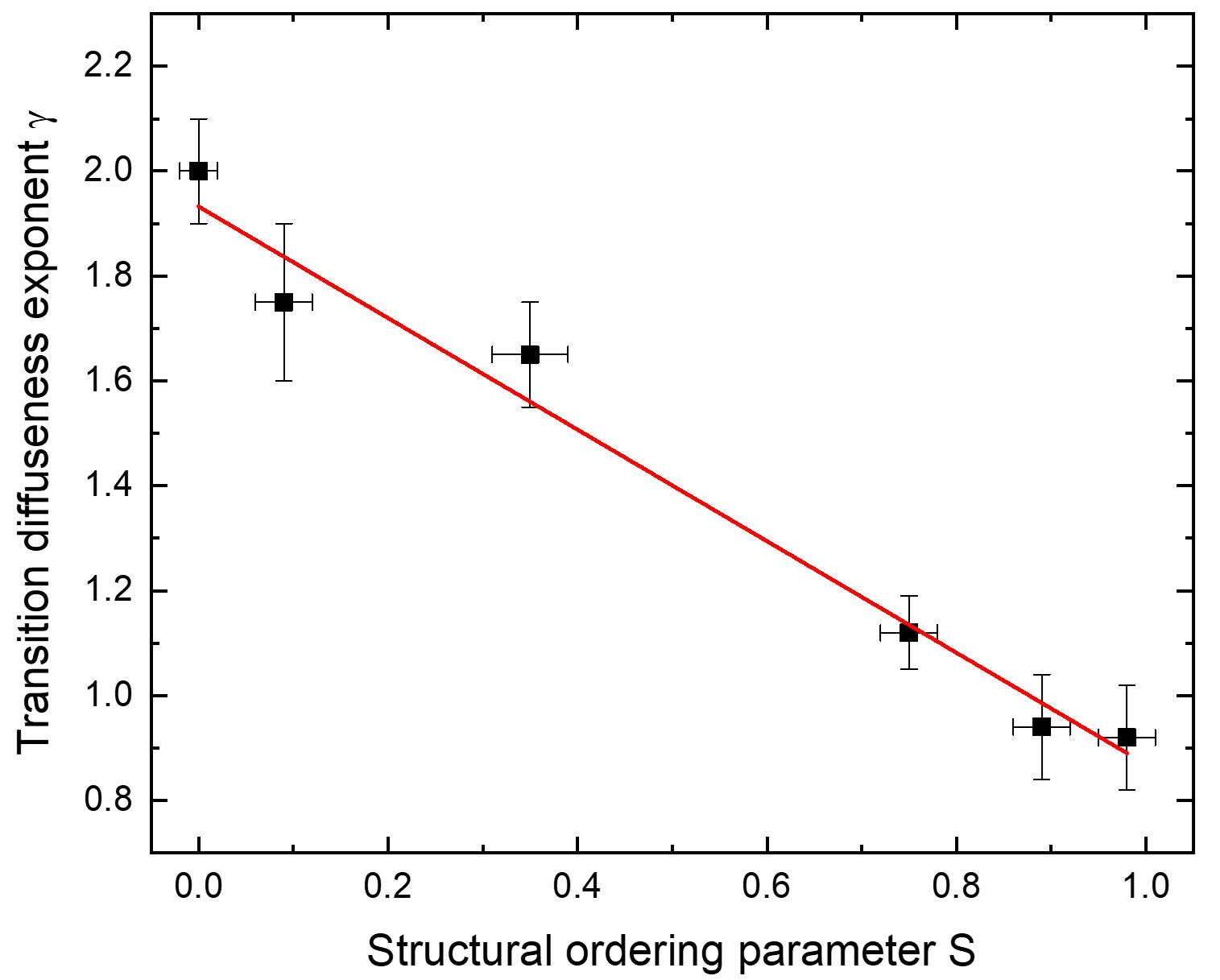}
\caption{\label{f4} Diffuseness exponent $\gamma$ as function of structural ordering parameter $S$ for all samples measured at 1\,kHz.}
\end{figure}

Figure \ref{f3} shows that the experimental data of all samples is well described by eq.\,\ref{Diffuse}. Both the shape of the curves in Fig.\,\ref{f3}(a) and the decreasing slope of the straight lines in Fig.\,\ref{f3}(b) already indicate that the diffuseness parameter $\gamma$ decreases with increasing degree of structural ordering. This is demonstrated even more clearly in Fig.\,\ref{f4}, which shows $\gamma$, determined at 1\,kHz, as function of the structural ordering parameter $S$. The values cover the entire range from $\gamma = 2$ for the disordered film with $S = 0$, indicating full relaxor behavior, to $\gamma = 0.92$ for the most highly ordered bulk sample with $S = 0.98$, agreeing with the expected value of $\gamma = 1$ for a ferroelectric phase transition within the error margin. The data can be approximated by the linear relation $\gamma(S) = 1.93 - 1.06*S$. The remarkable fact that this relation adequately describes the behavior of all samples - despite the otherwise very different dielectric properties of the bulk ceramics, MLCs and thin films - is an indication of the strength of the 'diffuseness exponent' $\gamma$ as an electrical measure to quantify relaxor behavior and disorder.

\section{Summary \& Conclusion}

While the relevance of electrical and structural disorder for relaxor ferroelectric behavior is well established, developing a quantitative correlation between disorder, relaxor character and diffuseness of phase transition has proven elusive, as modification of the degree of ordering is usually achieved by chemical variations that introduce additional variables to the experiment. Lead scandium tantalate offers the almost unique opportunity to study a ferroelectric system with varying degrees of structural (dis-)order without changes to the stoichiometry. Here, the degree of ordering was adjusted by thermal annealing treatments and experimentally analyzed by measuring the XRD (111) superstructure peak, resulting in the ordering parameter $S$. At the same time, the relaxor character of the phase transition was quantified through the diffuseness exponent $\gamma$, and a linear relation between the two quantities was established. This not only verifies earlier reports that the diffuseness parameter allows to compare the relaxor character of systems with strongly different values of $T_M$ and $\varepsilon_M'$ \cite{Baskaran01}, but also provides a quantitative relation between structural order and phase transition diffuseness that covers strongly different sample geometries and mechanical boundary conditions.

\section*{Acknowledgements}
We acknowledge the Fonds National de la Recherche (FNR) of Luxembourg for partly supporting this work through the project BRIDGES2021/MS/16282302/CECOHA/Defay. This work was also partly supported by the European Research Council (Project 101141445 — ELEC\_FROM\_HEAT — ERC 2023-ADG). H.U. and I.G. thank the Slovenian Research Agency (project J2-60035 and core funding P2-0105).


\end{document}